\documentclass[structabstract]{aa}
\usepackage{color}
\usepackage{graphicx}
\usepackage{txfonts}
\usepackage{lmodern}
\usepackage{natbib,twoopt}
\usepackage[breaklinks=true]{hyperref}
\usepackage{lscape}
\usepackage{multirow}
\usepackage{todonotes}

\hypersetup{colorlinks=true,linkcolor=blue,citecolor=blue,urlcolor=blue}

\definecolor{red}{rgb}{1,0,0}
\definecolor{blue}{rgb}{0,0,1}
\definecolor{mygreen}{RGB}{0,128,0}

\graphicspath{{pic/}}

\bibpunct{(}{)}{;}{a}{}{,} %% natbib format for A&A and ApJ
\makeatletter
\newcommandtwoopt{\citeads}[3][][]{\href{https://ui.adsabs.harvard.edu/\#abs/#3}%
{\def\hyper@linkstart##1##2{}%
\let\hyper@linkend\@empty\citealp[#1][#2]{#3}}}
\newcommandtwoopt{\citepads}[3][][]{\href{https://ui.adsabs.harvard.edu/\#abs/#3}%
{\def\hyper@linkstart##1##2{}%
\let\hyper@linkend\@empty\citep[#1][#2]{#3}}}
\newcommandtwoopt{\citetads}[3][][]{\href{https://ui.adsabs.harvard.edu/\#abs/#3}%
{\def\hyper@linkstart##1##2{}%
\let\hyper@linkend\@empty\citet[#1][#2]{#3}}}
\newcommandtwoopt{\citeyearads}[3][][]%
{\href{https://ui.adsabs.harvard.edu/\#abs/#3}
{\def\hyper@linkstart##1##2{}%
\let\hyper@linkend\@empty\citeyear[#1][#2]{#3}}}
\makeatother

\begin{document}
        
\title{The convective photosphere of the red supergiant \object{CE Tau}
\thanks{Based on observations collected at the European Organisation for Astronomical Research in the Southern Hemisphere under ESO programs 298.D-5005(A) and 298.D-5005(B).}\fnmsep
\thanks{The images from Fig. \ref{Fig:recons_mean} and \ref{Fig:image_recons_std} are available in electronic format the CDS via anonymous ftp to \texttt{cdsarc.u-strasbg.fr} (130.79.128.5) or via \url{http://cdsweb.u-strasbg.fr/cgi-bin/qcat?J/A+A/}}
}
\subtitle{I. VLTI/PIONIER H-band interferometric imaging}

\author{
        M.~Montarg\`es\inst{1, 2}
        \and
        R.~Norris\inst{3}
        \and
        A.~Chiavassa\inst{4}
        \and
        B.~Tessore\inst{5}
        \and
        A. L\`ebre\inst{5}
        \and
        F.~Baron\inst{3}
}

\institute{
    Institute of Astronomy, KU Leuven, Celestijnenlaan 200D B2401, 3001 Leuven, Belgium
    \and
        Institut de Radioastronomie Millim\'etrique, 300 rue de la Piscine, 38406, Saint Martin d'H\`eres, France
        \and
        Center for High Angular Resolution Astronomy, Georgia State University
        \and
        Universit\'e C\^ote d'Azur, Observatoire de la C\^ote d'Azur, CNRS, Lagrange, CS 34229, Nice, France
        \and
        Universit\'e de Montpellier, CNRS, LUPM, Place E. Bataillon, 34090, Montpellier, France 
}

\date{Received 29 June 2017; Accepted 03 February 2018}

\abstract
%Context heading
{Red supergiant stars are one of the latest stages in the evolution of massive stars. Their photospheric convection may play an important role in the launching mechanism of their mass loss; however,  its characteristics and dynamics are still poorly constrained.}
%Aims heading
{By observing red supergiant stars with near infrared interferometry at different epochs, we expect to reveal the evolution of bright convective features on their stellar surface.}
%Methods heading
{We observed the M2Iab-Ib red supergiant star CE Tau with the VLTI/PIONIER instrument in the H band at two different epochs separated by one month.}
%Results heading
{We derive the angular diameter of the star and basic stellar parameters, and reconstruct two reliable images of its H-band photosphere. The contrast of the convective pattern of the reconstructed images is $5 \pm 1 \%$ and $6 \pm 1 \%$ for our two epochs of observation.}
%Conclusions heading
{The stellar photosphere shows few changes between the two epochs. The contrast of the convective pattern is below the average contrast variations obtained on 30 randomly chosen snapshots of the best matching 3D radiative hydrodynamics simulation: $23 \pm 1~\%$ for the original simulation images and $16 \pm 1~\%$ for the maps degraded to the reconstruction resolution. We offer two hypotheses to explain this observation. CE Tau may be experiencing a quiet convective activity episode or it could be a consequence of its warmer effective temperature (hence its smaller radius) compared to the simulation.}

\keywords{Stars: individual: CE Tau; Stars: imaging; Stars: supergiants; Stars: mass-loss; Infrared: Stars, Techniques: interferometric}

\maketitle

%__________________________________Introduction
\section{Introduction \label{Sect:Intro}}

Most of the chemical elements in the Universe were forged inside evolved stars. As one of the latest stages in the evolution of massive stars, red supergiant (RSG) stars contribute to this enrichment through their mass loss. The mechanism that launches the material away from the star remains unknown. One scenario involves convection. From spectroscopic observations, \citetads{2007A&A...469..671J} suggested that by lowering the effective gravity, the turbulent velocity field associated with convection allows the radiative pressure on molecular lines to start the outflow. \citetads{1975ApJ...195..137S} predicted that photospheric convection on RSG stars would be different from what is known on solar-type stars: only a handful of giant granules would be present on the stellar surface. Near-infrared (NIR) observations by \citetads{2009A&A...508..923H} on \object{Betelgeuse}, the prototypical M-type RSG star, showed that only one or two bright and large hot spots were present on the photosphere. These features were interpreted as the top of convective granules. Additional observations were recently obtained that point toward a convection-based launch mechanism: ALMA observations show a large bright spot on the photosphere of Betelgeuse \citepads{2017A&A...602L..10O}. Its position matches the direction of a strong linearly polarized clump observed with VLT/SPHERE at three stellar radii \citepads{2016A&A...585A..28K}  that has been interpreted as recently formed dust. \citetads{2018A&A...609A..67K}  determined that the rotation axis of the star is also aligned with these two features and suggested that enhanced mass loss was emitted from the polar region due to the long term presence of a ``rogue'' convective cell. However, the velocity field derived by \citetads{2017Natur.548..310O} on the RSG Antares leads them to the conclusion that convection alone is not able to explain the atmospheric extension and motions of this star. A similar conclusion was obtained by \citetads{2015A&A...575A..50A} on a sample of RSG stars.

Furthermore, \citetads{2011A&A...528A.120C} showed that these large granules can additionally cause photocenter displacements significant enough to bias parallax measurements. As RSG stars can be used as bright candles at large distances, this effect has important consequences.

\citetads{2016A&A...588A.130M} monitored the H-band photosphere of Betelgeuse between January 2012 and November 2014. They obtained four epochs of observations showing a large feature (characteristic size $\sim$ R$_\star$) departing from spherically symmetric limb-darkened disk (LDD) models. These observations were correlated with January and November 2014 spectropolarimetric measurements at optical wavelengths obtained at the Narval instrument mounted on the Telescope Bernard Lyot (TBL) at the Pic du Midi observatory. From these observations, \citetads{2016A&A...591A.119A} discovered the linearly polarized spectrum of Betelgeuse. Their  analysis points to a continuum depolarization of Betelgeuse (due to scattering at photospheric level) that may be  related to brightness inhomogeneities lying at the stellar surface. The location and evolution of these inhomogeneities were mapped with an analytic model.

We report here NIR interferometric observations conducted on the RSG star \object{CE Tau} (119 Tau, HR 1845, HD 36389). CE Tau has a M2Iab-Ib spectral type. It has no reported companion. \citetads{2015MNRAS.446.3277C} observed significant departure from centrosymmetry on this star using the AMBER instrument at the Very Large Telescope Interferometer (VLTI). In Sect. \ref{Sect:Observations} we present the observations we obtained with VLTI/PIONIER. The angular diameter of the star is determined in Sect. \ref{Sect:Diameter}. Photospheric features are studied in Sect. \ref{Sect:Features} using several approaches: classical spotty models, image reconstruction, and 3D radiative hydrodynamics simulations. The updated angular diameter value is used to derive updated stellar parameters. The contrast of the convective pattern is discussed in Sect. \ref{Sect:Discussion}. We present a summary and conclusions in Sect. \ref{Sect:Conclusion}.
% A contemporaneous spectropolarimetric dataset was obtained during a large program on the TBL/Narval instrument and will be fully presented in a dedicated paper (Tessore et al. in prep.)
%__________________________________Observations
\section{Observations and data reduction\label{Sect:Observations}}

\begin{figure*}
        \centering
        \resizebox{\hsize}{!}{\includegraphics{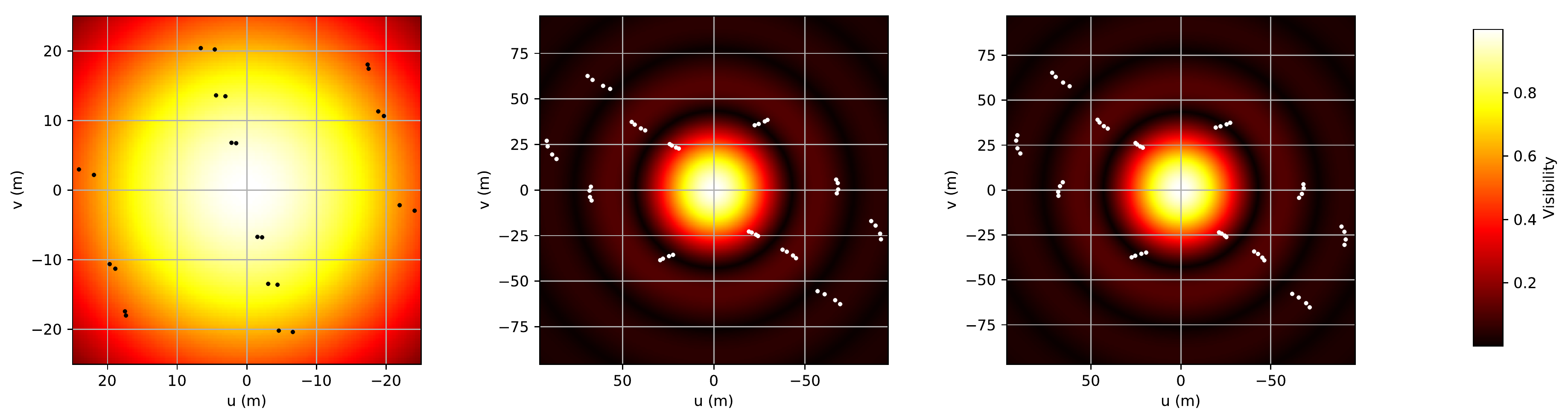}}
        \caption{($u, v$) coverage of our three epochs of VLTI/PIONIER observations of CE Tau. \textit{Left:} 14 November 2016. \textit{Center:} 22 November 2016. \textit{Right:} 23 December 2016. The underlying image corresponds to the best fit LDD model of Sect. \ref{Sect:Diameter}\label{Fig:uv_cov}}
\end{figure*}

CE Tau was observed on  14 and 22 November and on 22 and 23 December 2016 at the European Southern Observatory's VLTI \citepads{2010SPIE.7734E...3H} located on top of Cerro Paranal in Chile. We used the Precision Integrated-Optics Near-infrared Imaging ExpeRiment   (PIONIER, \citeads{2011A&A...535A..67L}) instrument, equipped with the RAPID detector, fed by the four 1.8m diameter Auxiliary Telescopes (AT) in their compact (stations A0-B2-C1-D0) and intermediate (stations D0-G2-J3-K0) configurations. Ground baselines were between 6.9 and 96.9~m. The GRISM was set in the optical path of PIONIER, providing its highest available spectral resolution of R $\sim 30$, delivering six spectral channels between 1.51 and 1.77~$\mu$m. The log of the observations is available in Table \ref{Tab:ObsLog}. We note that other executions of the observing blocks exist in the archive, but they had a low quality grade. The data from 22 December show some visibility loss. As the observation was repeated on 23 December without any issue, and because we do not expect the photosphere of a RSG star to evolve significantly
over 24h, we will not consider the 22 December data  further. The ($u, v$) coverage for each of the three epochs we consider is represented  in Fig. \ref{Fig:uv_cov}.

The data were reduced and calibrated with the publicly available PIONIER pipeline \citepads{2011A&A...535A..67L} version 3.51. The angular diameters of our interferometric calibrators are listed in Table \ref{Tab:calibrators_data}. The uncertainties on the observables are directly computed by the pipeline: on the uncalibrated data it derives the statistical dispersion over 100 scans of each $\sim 30$~s exposure. Then for the calibrated product it quadratically adds the error from the transfer function. Each individual observation results in six squared visibilities and four closure phases per spectral channel.

\begin{table}
        \caption{Adopted uniform disk diameters for the interferometric calibrators.}
        \label{Tab:calibrators_data}
        \centering
        \begin{tabular}{ll}
                \hline \hline
                \noalign{\smallskip}
                Name & Diameter (mas) \\
                \hline
                \noalign{\smallskip}
                \object{HR 1684} & $ 2.60 \pm 0.03$ \\
                \object{$\phi$02 Ori} & $2.13 \pm 0.02$ \\      
                \hline
        \end{tabular}
        \tablebib{\citetads{2002A&A...393..183B}}
\end{table}

%__________________________________Analytical model analysis
\section{Angular diameter measurements\label{Sect:Diameter}}

\begin{figure}
        \centering
        \resizebox{\hsize}{!}{\includegraphics{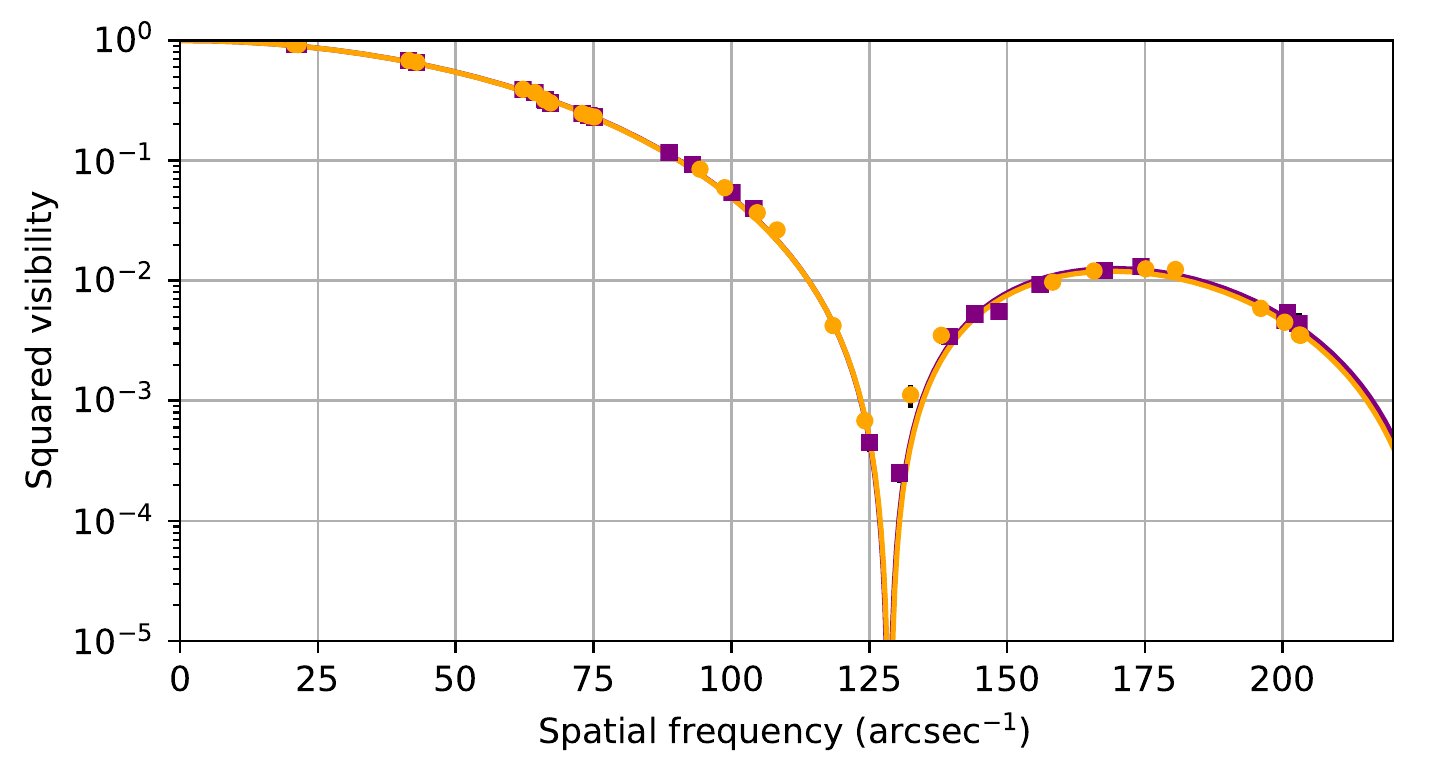}}
        \caption{Squared visibility in the first two lobes measured by VLTI/PIONIER on CE Tau at 1.62~$\mu$m. The purple squares correspond to the November dataset. The orange dots correspond  to the December dataset. The continuous curves are the best fit LDD power-law models for both epochs.\label{Fig:fit_LDD}}
\end{figure}

Our data from 14 November 2016 were acquired in the compact configuration. These observations cover only the first lobe of the visibility function (Fig. \ref{Fig:uv_cov}, left). Data from 22 November and 23 December cover the first three lobes (Fig. \ref{Fig:uv_cov}, center and right). When plotting the visibility versus the spatial frequency (Fig.~\ref{Fig:fit_LDD}), we observe that the low spatial frequency first lobe data between 22 November and 23 December are invariant (while the signal changes for the longer baselines). Lower spatial frequencies probe larger scale features and in the case of the first lobe, the general stellar shape. These are crucial for reliable model fitting and image reconstruction. Therefore, we decided to merge the 14 November data on the compact configuration (very short baselines) with the intermediate configuration data of 22 November (November dataset hereafter) and with the 23 December data (December dataset hereafter). This allowed us to have short baseline data on both epochs.\\

To determine the angular diameter of the star, we used both a uniform disk (UD, $I = I_0$) and a LDD power-law model ($I/I_0 = \mu^\alpha$). The visibility amplitude of the latter is given by \citepads{1997A&A...327..199H}
\begin{equation}
        V_\mathrm{LDD}(s) = \Gamma(\nu+1)\frac{J_\nu (x)}{(x/2)^\nu}\label{Eq:VisLDD}
,\end{equation}
\noindent where $\nu = \alpha/2 +1$, $s$ is the spatial frequency, $x=\pi s \theta_\mathrm{LDD}$, $\theta_\mathrm{LDD}$ the LDD angular diameter, $J_\nu$ is the first species Bessel function of order $\nu$, and $\Gamma$ is the Euler function.

In order to derive the angular diameter, we consider here only the 1.62~$\mu$m channel, closest to the H$^-$ opacity minimum \citepads{2008AJ....135.1450G}. A global fit showed important deviations in the third lobe, probably due to contaminations by inhomogeneities. To avoid a bias in the diameter estimation, we present only the fit restricted to the first and second lobes (spatial frequencies lower than 220~arcsec$^{-1}$). The results of the fit are given in Table \ref{Tab:fit_LDD} and the LDD model is represented in Fig. \ref{Fig:fit_LDD} with the data. The UD model reproduces the data poorly and will not be considered further.

\begin{table}
        \caption{Best fit parameters for UD and power-law LDD at 1.62~$\mu$m for the first two lobes of the squared visibility function.}
        \label{Tab:fit_LDD}
        \centering
        \begin{tabular}{lll}
                \hline \hline
                \noalign{\smallskip}
                Parameter & Nov. values & Dec. values\\
                \hline
                \noalign{\smallskip}
                $\theta_\mathrm{UD}$ (mas) & $ 9.61 \pm 0.26$ & $ 9.91 \pm 0.28$\\
                $\tilde{\chi}^2_\mathrm{UD}$ & 89.8 & 91.2 \\
                \noalign{\smallskip}
                $\theta_{\mathrm{LDD}}$ (mas) & $ 10.09 \pm 0.09$ & $ 10.18 \pm 0.07$ \\
                $\alpha_{\mathrm{LDD}}$ & $0.36^{+0.13}_{-0.05}$ & $0.43 \pm 0.07$ \\
                $\tilde{\chi}^2_\mathrm{LDD}$ & 6.35 & 6.85 \\
                \hline
        \end{tabular}
        \tablefoot{$\tilde{\chi^2}$ corresponds to the reduced $\chi^2$. The uncertainties on the fitted values are derived with the parameter values for which $\tilde{\chi^2} = 2\tilde{\chi}^2_\mathrm{min}$.}     
\end{table}

The fits of the two epochs of observations by the LDD model give different but compatible values. For a fair comparison with the models including inhomogeneities presented in Sect. \ref{Sect:Features}, we also derived the reduced $\chi^2$ for the best LDD model taking into account both the visibilities and the closure phases of the entire spatial frequency domain. We obtain $\tilde{\chi}^2 = 3106$ and 4578 for the November and December datasets, respectively. We do not fit the LDD model to the closure phases as this observable  constrains disk models poorly (our dataset does not sample  the flips from 0 to $\pi$ exactly). The most recent angular diameter measurement of this star is given by \citetads{2013MNRAS.434..437C} with 9.97~mas in the K band measured with VLTI/AMBER on November 2009. These authors compiled all the previous angular diameter measurements of this star. For different spectral domains the values range between 9.4 and 13.0~mas for indirect methods, 9.1 and 17~mas for lunar occultations, and 9.3 and 10.68~mas for long baseline interferometry. Therefore, with values of $10.09 \pm 0.09$~mas and $10.18 \pm 0.07$~mas, our LDD diameters appear compatible with these results. The poor reduced $\chi^2$ of the LDD model when compared to the closure phase, although the squared visibility value is acceptable, suggests the presence of photospheric structures that cannot be reproduced by simple disk models.

\section{Photospheric features\label{Sect:Features}}

\subsection{Limb-darkened disk and Gaussian spots\label{Sect:spots}}

\subsubsection{Model with one spot}

The presence of inhomogeneities on the photosphere of RSG stars is expected. Previous observations \citep[e.g.,][]{2009A&A...508..923H,2010A&A...515A..12C,2014ApJ...785...46B,2016A&A...588A.130M,2016A&A...591A.119A,2017Natur.548..310O,2017A&A...606L...1W} have shown that they are common. They are also predicted by models \citepads{1975ApJ...195..137S,2011A&A...535A..22C}. The most economical models for inhomogeneities are spots, either uniform disk or Gaussian. Here, following the models used by \citetads{2016A&A...588A.130M,2017A&A...605A.108M}, we  use Gaussian spots. We consider the LDD model presented in Sect. \ref{Sect:Diameter}, and we add a Gaussian spot at the position ($x_\mathrm{center}, y_\mathrm{center}$) relative to the center of the stellar disk.  We denote w$_\mathrm{LDD}$ and w$_\mathrm{spot}$ the peak flux of the LDD model and the spot, respectively, and FWHM is its full width at half
maximum. We normalize the model using

\begin{equation}
        w_\mathrm{LDD} + w_\mathrm{spot} = 1. 
\end{equation}
        
\noindent The complex visibility of the model is then

\begin{equation}
        V_\mathrm{model} = w_\mathrm{LDD} V_\mathrm{LDD} + w_\mathrm{spot} V_\mathrm{spot},
\end{equation}

\noindent with

\begin{equation}
        \begin{array}{l}
                V_\mathrm{spot}(u, v) = \exp\left[-\frac{(2 \pi f \sigma)^2}{2}\right] \\ \\ \times \exp\left[-2i\pi(ux_\mathrm{center} + vy_\mathrm{center})\right]
        \end{array}
,\end{equation} 

\noindent where $i^2 = -1$, $f = \sqrt{u^2 + v^2}$ and $\sigma = FWHM/(2\sqrt{2\ln(2)})$.

Modeling such features is difficult as the $\chi^2$ distribution becomes very complex, as discussed by \citetads{2014ApJ...785...46B}. Therefore, the fitting process is done in two steps for each epoch. In the first step, the parameters of the limb-darkened disk are fixed to the values derived in Sect. \ref{Sect:Diameter}. The FWHM and $w_\mathrm{spot}$ explore the ranges [0.05; 5 mas] by steps of 0.5 mas and [-0.5; 0.5] by steps of 0.05, respectively. For each couple (FWHM, $w_\mathrm{spot}$) a $\chi^2$ map is built by deriving the reduced $\chi^2$ associated with the model for various positions of the Gaussian spot on the stellar disk. We used 50 by 50 pixels maps whose edges correspond to the stellar radius. The maps obtained for the best couple (FWHM, $w_\mathrm{spot}$) are shown in Fig. \ref{Fig:chi2map_1spot}. The parameters associated with the minimum $\chi^2$ over this set of maps are used as initial guesses in the second step of the fitting process. In this next step, a Levenberg--Marquardt fit is done on all parameters. The results are presented in Table \ref{Tab:fit_1spot}. We also derived the F parameter :

\begin{equation}
         F = \frac{\chi^2_\mathrm{LDD} - \chi^2_\mathrm{1 spot}}{N_\mathrm{param, 1spot} - N_\mathrm{param, LDD}} \times \frac{N_\mathrm{data} - N_\mathrm{param, 1spot}}{\chi^2_\mathrm{1spot}}.
\end{equation}

\noindent This parameter allows us to determine the significance of a fit with a large number of parameters. For a (2, 6) distribution and a significance of 5\%, the F parameter must be above 5.143\footnote{\url{http://www.itl.nist.gov/div898/handbook/eda/section3/eda3673.htm}}. In principle, the F parameter should be derived separately for the squared visibilities and the closure phases, but \citetads{2017A&A...605A.108M} have shown that it is irrelevant for the bright targets accessible to optical interferometers.

\begin{table}
        \caption{Best fit parameters for a single Gaussian spot on a LDD model for  the squared visibilities and for the closure phases.}
        \label{Tab:fit_1spot}
        \centering
        \begin{tabular}{lll}
                \hline \hline
                \noalign{\smallskip}
                Parameter & Nov. values & Dec. values\\
                \hline
                \noalign{\smallskip}
                $\theta_{\mathrm{LDD}}$ (mas) & $ 9.94 \pm 0.03$ & $ 10.04 \pm 0.03$ \\            $\alpha_{\mathrm{LDD}}$ & $0.34 \pm 0.02$ & $0.39 \pm 0.02 $ \\
                $w_\mathrm{spot}$ & $0.04 \pm 0.01$ & $0.04 \pm 0.01$ \\
                $x_\mathrm{center}$ (mas) & $-0.57 \pm 0.14$ & $-0.50 \pm 0.16$ \\
                $y_\mathrm{center}$ (mas) & $2.84 \pm 0.15$ & $2.81 \pm 0.24$ \\
                $FWHM$ (mas) & $4.41 \pm 0.29$ & $3.87 \pm 0.33$ \\
                $\tilde{\chi}^2_\mathrm{LDD}$ & 9.1 & 13.6 \\
                $F$ & 4860 & 4851 \\
                \hline
        \end{tabular}
\end{table}

The very high values of the F parameter indicate that adding a single Gaussian spot strongly improves the fitting process without overfitting. We note that with our approach, bright and dark spots were allowed (positive or negative flux), but the model converged to a bright spot, with very nearby parameters for both epochs.

Attempts were made to fit a two-spot model, but no convergence was reached for the December dataset. Our conclusion is that the $\chi^2$ distribution for a two-spot model is much more complex: there is an important correlation between the characteristics of the two spots, making the determination of the absolute maximum an extremely complex task. Therefore, we limit our analysis to the single-spot case.

\subsubsection{Basic stellar parameters\label{Sect:Stellar_params}}

From the determination of the LDD angular diameter, we are able to derive several basic characteristics of the star. With the parallax of $1.82 \pm 0.26$~mas \citepads{2007A&A...474..653V}, we are able to derive the linear radius of the star $R = 587 \pm 85$~R$_\odot$ in November, and $R = 593 \pm 86$~R$_\odot$ in December. These values are in agreement with the value of $601 \pm 83$~R$_\odot$ from \citetads{2013MNRAS.434..437C}.

To determine the bolometric flux, we can use photometric measurements from \citetads{1970ApJ...162..217L} and \citetads{2002yCat.2237....0D} ranging from the U to the N band (0.36 to 10.2~$\mu$m). After determining the difference between the B-V of these data (2.08) and the intrinsic B-V color of a M2Iab-Ib star (1.69) reported in \citetads{1985ApJS...57...91E}, we can compute the interstellar extinction A$_\lambda$ in each filter using the data in \citet{1979ARA&A..17...73S}. From this we derive the bolometric flux F$_\mathrm{UBVRIJHKLN} = 7.01 \times 10^{-9}$~W.m$^{-2}$. Because CE Tau  is a semi-regular variable, we use the 1 magnitude visual dispersion reported by the American Association of Variable Star Observers (AAVSO) to adopt a 12.8\% uncertainty on the bolometric flux \citepads{2013A&A...555A..24O}: $\pm 8.98 \times 10^{-10}$~W.m$^{-2}$.

The LDD angular diameter and the bolometric flux give us access to the effective temperature: T$_\mathrm{eff} = 3820 \pm 135$~K in November and T$_\mathrm{eff} = 3801 \pm 134$~K in December. The 3$\sigma$ range of these values largely encompasses the value of 3660~K given for a M2 RSG by \citetads{2005ApJ...628..973L} and the 3700~K of \citetads{1980ApJ...241..218L}. We can also use the parallax and  the bolometric flux to derive the luminosity of the star: $\log L/L_\odot = 4.82^{+0.12}_{-0.16}$, a value compatible with that found by \citetads{2013MNRAS.434..437C}, who derived a value of 4.63 with an uncertainty of 13\%.

\begin{figure}
        \centering
        \resizebox{\hsize}{!}{\includegraphics{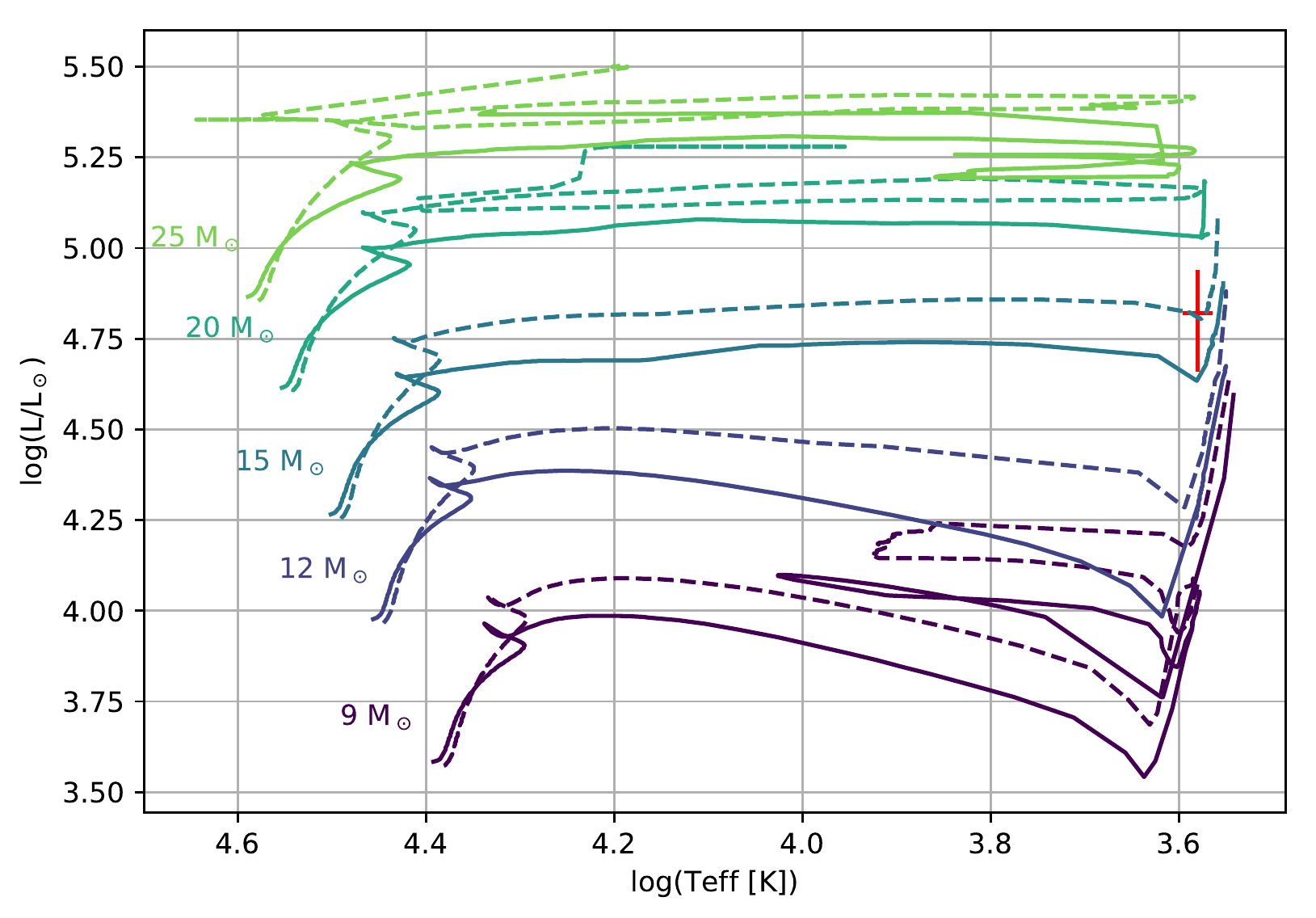}}
        \caption{Hertzsprung--Russell  diagram with the position of CE Tau marked by the red cross. The continuous (resp. dashed) lines correspond to the non-rotational (resp. rotational) evolutionary tracks of \citetads{2012A&A...537A.146E} for solar metallicity abundances (Z=0.014).\label{Fig:evolution}}
\end{figure}

Placing CE Tau in a Hertzsprung--Russell (H-R) diagram  (Fig. \ref{Fig:evolution}), we see that the derived values agree remarkably well with the evolutionary track \citepads{2012A&A...537A.146E} of a 15~M$_\odot$ star with rotation (the rotation rate on the zero age main sequence is v$_\mathrm{ini}$/v$_\mathrm{crit}$ = 0.4) and solar metallicity (Z=0.014). According to \citetads{1980ApJ...241..218L} the metallicity of CE Tau is [Fe/H]~=~0.11. Given the error bars on the parameters we adopt an uncertainty of 2~M$_\odot$ on the stellar mass. For an initial mass of 15~M$_\odot$, the evolutionary models predict that at its current age, CE Tau weighs M$_\mathrm{cur} = 14.37^{+2.00}_{-2.77}$~M$_\odot$.  With this mass and the linear radius, we can derive the surface gravity: $\log g = 0.05^{+0.11}_{-0.17}$, a value compatible with $\log g = 0.07$ obtained by \citetads{1980ApJ...241..218L}. We can also use the location of CE Tau on the H-R diagram and the evolutionary models (with and without rotation) to estimate the age of the star: $13.9^{+1.0}_{-2.5}$~Myr.

%With the linear radius, we can to derive the surface gravity: $\log g = 0.05^{+0.11}_{-0.17}$, with $g$ in units of cm.s$^{-2}$. This value is compatible with the 0.07 obtained by \citetads{1980ApJ...241..218L}.
%
%The location of CE Tau on the H-R diagram and the evolutionary models (with and without rotation) allow us to estimate the age of the star: $13.9^{+1.0}_{-2.5}$~Myr.

\subsection{Image reconstructions\label{Sect:Image}}

As fitting analytical models to reproduce photospheric inhomogeneities does not give unequivocal results due to the complexity of the $\chi^2$ distribution, we decided to continue our analysis with the image reconstruction technique. Our ($u, v$) coverage actually allows this procedure if we take into account the synthesized beam, while our limited resolution relative to the angular diameter of the star does not enable us to use the statistical approach of \citetads{2017A&A...605A.108M}.

To produce images from the datasets, we use \texttt{SQUEEZE}, a compressed sensing-based image reconstruction tool. \texttt{SQUEEZE} uses a Markov chain Monte Carlo  (MCMC) approach to search the imaging probability space via parallel tempering \citepads{2010SPIE.7734E..2IB}. This method allows the simultaneous use of a variety of regularizers, including the l0 norm
 non-convex regularizer.

\begin{figure}
        \centering
        \resizebox{0.95\hsize}{!}{\includegraphics{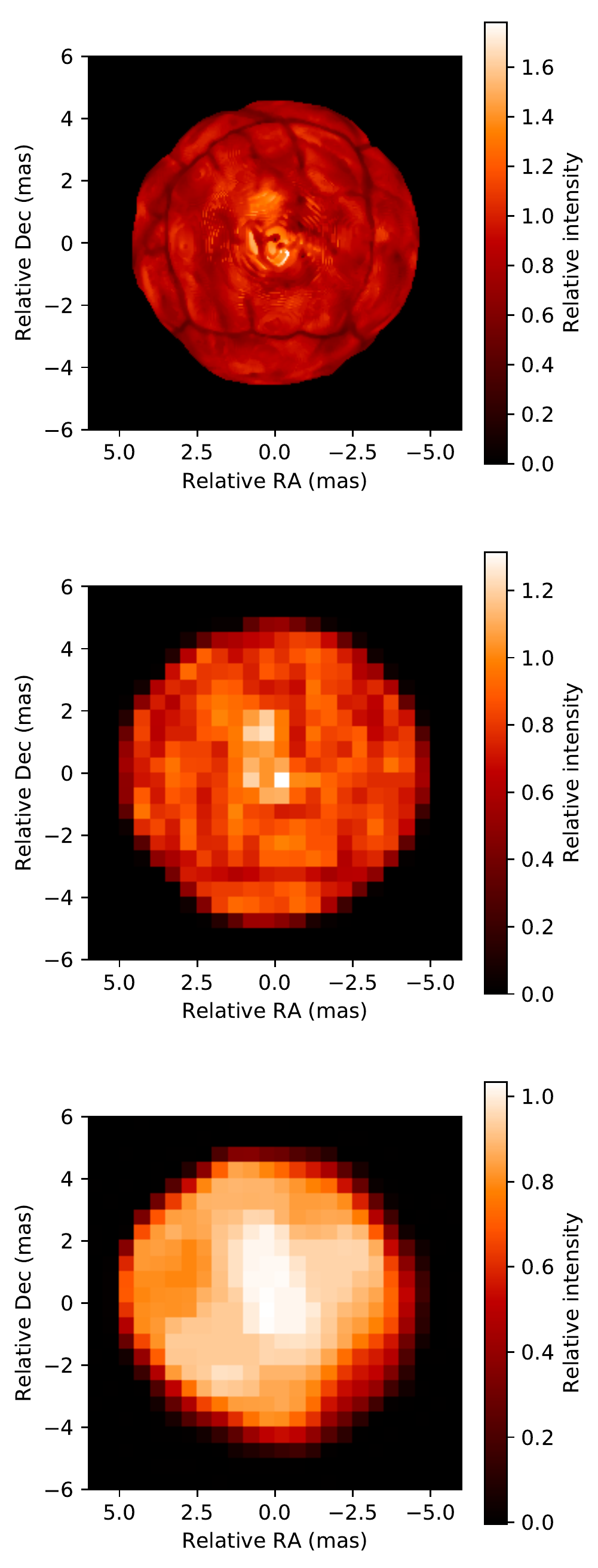}}
        \caption{Best snapshot of the RHD simulation. \textit{Top}: Intensity image. \textit{Center}: Intensity image degraded to the resolution of the \texttt{SQUEEZE} reconstructed images. \textit{Bottom}: Reconstructed image from a set of interferometric observables matching the ($u, v$) coverage and noise level of our PIONIER data. \label{Fig:Simus_profiles}}
\end{figure}

Because of the limited ($u, v$) coverage of our observations, we decided to use all the spectral channels at once and to start our reconstruction process by determining the best regularizers and regularization strengths for our data. We did this by using the 3D radiative hydrodynamics (RHD) model described in Sect. \ref{Sect:Simu} as the source image for a simulated observation, as this allowed us to compare reconstruction results to a known source (center and bottom images of Fig. \ref{Fig:Simus_profiles}). We produce the simulated observation using OIFITS-SIM\footnote{\url{https://github.com/fabienbaron/oifits-sim}}, copying the ($u, v$) coverage and noise statistics of our December observations. The reconstructions use masks and initial images based on uniform disks of sizes 10--12 mas, depending on the pixel scale. In order to correct for artifacts in the image reconstruction process, we ran five MCMC chains, each consisting of 500 iterations, in the end producing an average and error image after coalligning the mean result of each chain using the subpixel registration algorithm of \citet{Guizar-Sicairos:08}. In order to test for possible super-resolution in the reconstruction process, we used this method for resolutions of 2.0 mas, 1.0 mas, 0.8 mas, and 0.5 mas. We compared the resulting reconstructions to the source image, convolved to a resolution corresponding to the reconstruction: we coalligned the reconstruction to match the position of the convolved source image and used the $l1$-norm as our metric for comparison, which \citetads{2017MNRAS.465.3823G} found to be the best metric for assessing the quality of a reconstructed image.

Our comparison found that the best reconstructed images came from the reconstructions at 0.5 mas resolution, which is a significantly higher resolution than   expected from the maximum projected baseline of the observations. Thus, we opted to use the regularizers and strengths of that reconstruction, but also to reconstruct using the parameters of the next best reconstruction at a lower resolution, in this case at 1.0 mas. For the 0.5 mas resolution, we ran \texttt{SQUEEZE} using the same parameters as the best reconstruction of the simulated data: a 32x32 pixel grid, using both total variation and Laplacian regularizers, a mask of a 11 mas diameter uniform disk, an initial image of a 10.5 mas diameter uniform disk, 2000 elements, and 500 iterations. However, in this case we use 25 chains in order to better account for artifacts due to the reconstruction process. Once more we produce a single average and standard deviation image from the result of each chain. Reduced $\chi^{2}=1.53$ for the November dataset and $\chi^{2}=2.08$ for the December image are obtained. We follow the same procedure for the 1.0 mas reconstruction, the differences being that our image was 16x16 pixels and that we used total variation and the $l1$-norm of the \textit{\`{a} trous} wavelet transform \citepads{1989wtfm.conf..286H}. We find the reduced $\chi^{2}=1.78$ for November and $\chi^{2}=2.21$ for December.

\begin{figure}
        \centering
        \resizebox{\hsize}{!}{\includegraphics{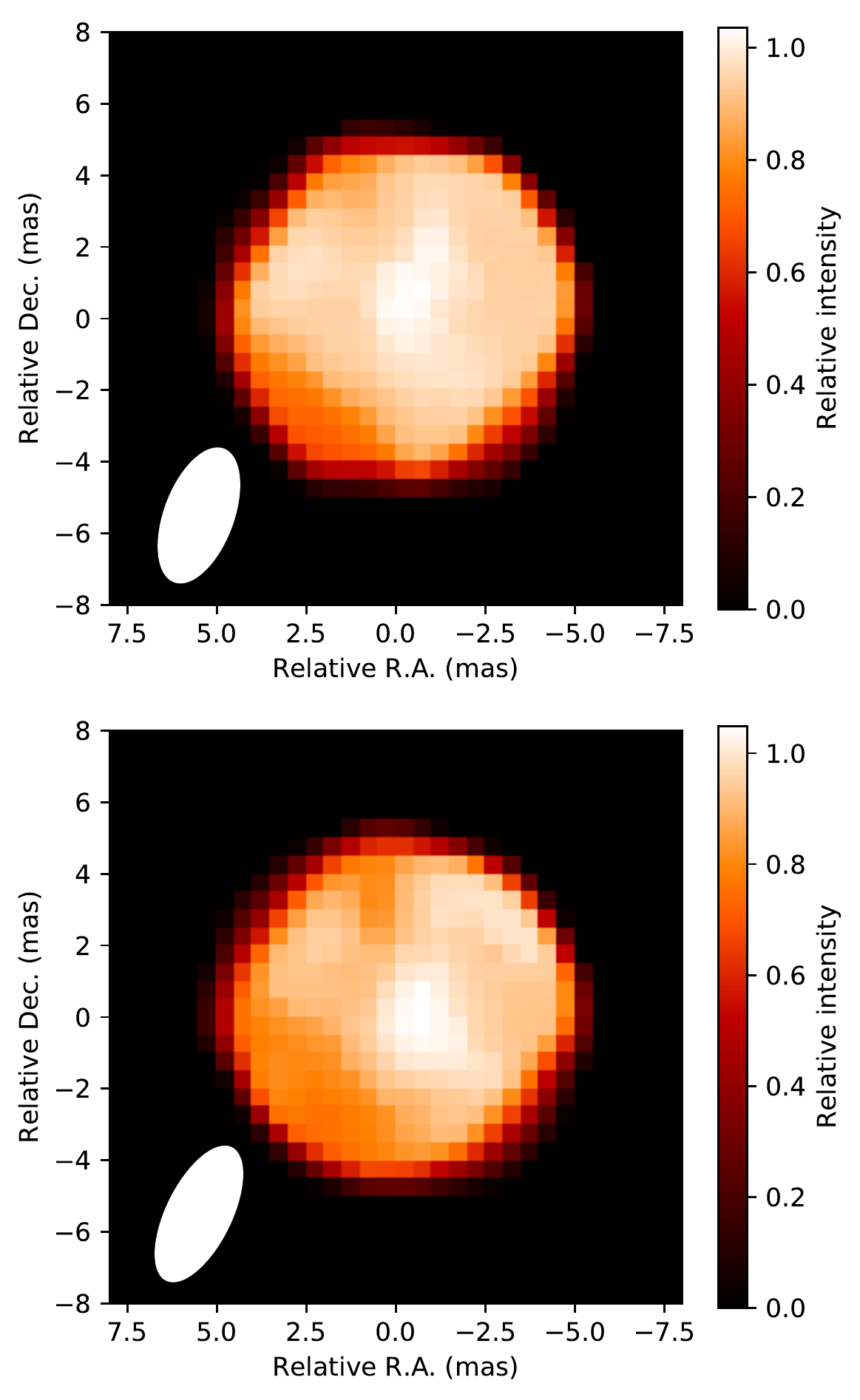}}
        \caption{Mean \texttt{SQUEEZE} image reconstruction of CE Tau for a pixel size of 0.5~mas. The top (bottom, resp.) image corresponds to the November (December, resp.) dataset. North is up and east is left. The white ellipse in the bottom left corner represents the main lobe of the synthesized beam. \label{Fig:recons_mean}}
\end{figure}

In order to assess the reliability of the reconstruction process, we also used the Multi-aperture Image Reconstruction Algorithm (\texttt{MIRA}) to reconstruct images \citepads{2008SPIE.7013E..43T}. In contrast to \texttt{SQUEEZE}, which uses MCMC minimization, \texttt{MIRA} uses gradient descent. To reconstruct our images, we used a 16x16 grid with a pixel scale of 1.0 mas/pixel and a 32x32 grid with a pixel scale of 0.5 mas/pixel, the same scales as used with \texttt{SQUEEZE}. We used the same initial images and masks as we did with \texttt{SQUEEZE}. Starting with the initial image, we ran \texttt{MIRA} with a smoothness regularizer for 200 evaluations, then used the resulting centered image as an initial image for a reconstruction using a maximum entropy regularizer with 500 evaluations. We then repeated the process using the resulting centered image as a starting point. In each case, we decreased the regularization strength, $\mu$ in each iteration for each regularizer. The resulting images from this method (Fig. \ref{Fig:Mira}) share characteristics, including position and size of bright features, with those obtained using \texttt{SQUEEZE}. From now on, we  only consider the 0.5~mas resolution \texttt{SQUEEZE} images for the analysis.

Figure \ref{Fig:recons_mean} represents the resulting mean images for the 0.5~mas resolution for both epochs. Differences are noticeable between the two epochs. As the ($u, v$) coverage is marginally different (Fig. \ref{Fig:uv_cov}), and considering the standard deviation of the images (Fig. \ref{Fig:image_recons_std}), we suggest that the differences come from a temporal evolution of the stellar surface and are not a residual of the imaging process, as predicted by \citetads{2011A&A...528A.120C}. The inhomogeneities are discussed further in Sect. \ref{Sect:Discussion}, and are compared with the best numerical model reproducing the observations.

\subsection{Comparison with 3D RHD simulations \label{Sect:Simu}}

Contrast variations of the NIR photosphere of RSG stars have been best explained by the presence of convective cells. Three-dimensional radiative hydrodynamics (RHD) simulations have been used to interpret the interferometric signals of several stars. For Betelgeuse, the comparison was successful in both the H and the K bands (\citeads{2010A&A...515A..12C} and \citeads{2014A&A...572A..17M}). However, more recent observations in the H band have shown that at different epochs on Betelgeuse \citepads{2016A&A...588A.130M} or at an unprecedented angular resolution on Antares \citepads{2017A&A...605A.108M}, these numerical models were unable to reproduce the departure from classical disk models. In the case of Betelgeuse, it could be the consequence of a change in the convective regime of the star \citepads{2018A&A...609A..67K} that may have lead to material ejection observed by VLT/SPHERE \citepads{2016A&A...585A..28K}. Because it is lacking from previous observations, it is impossible to conclude on Antares. Alternatively, we note that older observations were in agreement with the simulations: the new instrumentation available in interferometry (better angular resolution, four-telescope $u,v$ coverage) may bias the interpretation.

To assess the compatibility between our PIONIER data of CE Tau and the convective models, we used a numerical simulation produced by the COnservative COde for the COmputation of COmpressible COnvection in a BOx of L Dimensions, L = 2, 3 (CO$^5$BOLD, \citeads{2012JCoPh.231..919F}). We used the simulation st35gm03n13  \citepads{2011A&A...535A..22C} whose characteristics are presented in Table \ref{Tab:Charac_Simu}. This model does not include rotation or a magnetic field. \citetads{2017A&A...605A.108M} checked that at the 15th lobe of the visibility function, these simulations are not affected by numerical artifacts. We recall that our VLTI/PIONIER data on CE Tau probe only the first three lobes of the visibility function.

\begin{table*}
        \caption{Characteristics of the RHD simulation used to analyze our VLTI/PIONIER data. The stellar parameters of CE Tau were derived in Sect. \ref{Sect:Stellar_params}.}
        \label{Tab:Charac_Simu}
        \centering
        \begin{small}
                \begin{tabular}{ccccccccc}
                        \hline\hline
                        \noalign{\smallskip}
                        Model & M$_\star$ & L & T$_\mathrm{eff}$ & R$_\star$ & $\log g$ & Grid & Grid \\
                        & (M$_\odot$)& (L$_\odot$) & (K) & (R$_\odot$) & & (N & res. \\
                        & & & & & & points) & [$R_\odot$] \\
                        \hline
                        \noalign{\smallskip}
                        st35gm03n13 & 12 & $8.95 \pm 0.009 \times 10^4$ & $3430 \pm 8$ & $846.0 \pm 1.1$ & $-0.354 \pm 0.001$ & $235^3$ & 8.6 \\
                        CE Tau - Nov. & 14.37$^{+0.02}_{-1.91}$ & $6.61 ^{+2.10}_{-2.03} \times 10^4$ & $3820 \pm 135$ & $587 \pm 85$ & $0.05^{+0.11}_{-0.17}$& ... & ... \\
                        CE Tau - Dec. & 14.37$^{+0.02}_{-1.91}$ & $6.61 ^{+2.10}_{-2.03} \times 10^4$ & $3801 \pm 134$ & $593 \pm 86$ & $0.05^{+0.11}_{-0.17}$& ... & ... \\
                        \hline
                \end{tabular}
                \tablefoot{See \citeads{2011A&A...535A..22C} for more details.}
        \end{small}
\end{table*}

Hundreds of temporal snapshots are computed. Intensity images are computed in the PIONIER spectral channels using the 3D pure local thermodynamic equilibrium (LTE) radiative transfer code Optim3D \citepads{2009A&A...506.1351C}. The images are scaled to the angular diameter of CE Tau for each epoch. To account for the unknown orientation on the plane of the sky, each image is rotated around its center. We used 36 angle positions between 0$^\circ$ and 180$^\circ$. Finally, interferometric observables were derived using a fast Fourier transform (FFT) algorithm.

The observables associated with the grid of temporal snapshots and rotation angles were compared to the squared visibilities of our data. The aim of this procedure is not to find a snapshot reproducing  the configuration of the convective pattern of CE Tau exactly, but to find one that reproduces its characteristics (size of the cells, contrast). Matching a configuration pattern would require a much greater number of snapshots, due to the statistics of the convective configurations, and would not provide a better physical interpretation. For both epochs,  temporal snapshot \#091 was the best match, although the best orientation on the plane of the sky is changing. As shown by \citetads{2017A&A...605A.108M}, the angular diameter of the intensity images of the simulations has a strong impact on the quality of the match.  Therefore, for the best snapshot, we derive intensity images for each rotation angle at different angular diameters: from 9.59 to 10.59~mas for the November dataset and from 9.68 to 10.68~mas for the December dataset by steps of 0.05~mas. The best matches are obtained for an angular diameter of $10.24 \pm 0.05$~mas (resp. $9.98 \pm 0.05$~mas) and  a position angle of 0$^\circ$ for both epochs, and give a $\tilde{\chi}^2$ of 98.2 (resp. 118.6) for the November dataset (resp. December). Our observations of CE Tau are not well matched by 3D RHD simulations.

\section{Discussion \label{Sect:Discussion}}

\begin{figure}
        \centering
        \resizebox{\hsize}{!}{\includegraphics{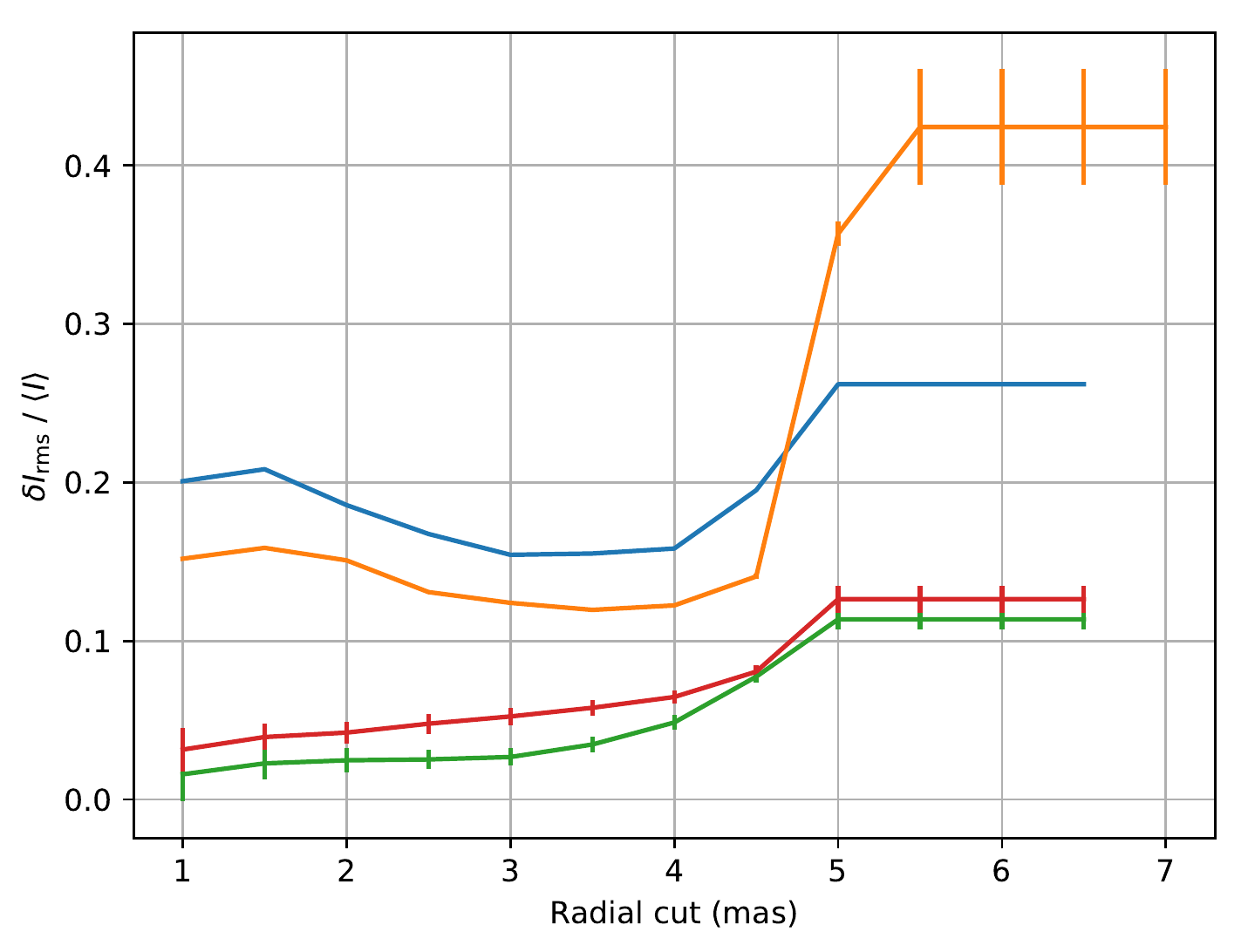}}
        \caption{Contrast of the convective pattern of the best convective simulation snapshot (orange), the same degraded to the reconstruction resolution (blue), the November (green), and December (red) reconstructed images as a function of the upper radius considered. \label{Fig:contrast_tremblay}}
\end{figure}

%\subsection{Contrast of the photospheric features \label{Sect:contrast}}
To better compare the \texttt{SQUEEZE} reconstructed images with the 3D RHD simulations, we use the method of \citetads{2017A&A...606L...1W} and derive the contrast $\delta I_\mathrm{rms}/\langle I \rangle$ defined in \citetads{2013A&A...557A...7T}. However, as we have only two epochs of observations and one simulation snapshot, we discard the temporal average and compare the elementary elements. Additionally, we correct for the limb-darkening effect by dividing each image by a best fit LDD intensity model. The resulting contrast is represented in Fig. \ref{Fig:contrast_tremblay} as a function of the upper radius considered. Near the limb of the star, this contrast is biased by the limb-darkening (LD) effect: small errors in the LD modeling can be interpreted as important fluctuations due to the low intensity at the edge of the stellar disk. Therefore, we will only consider the contrast below a radial cut of 4~mas. We also impose a lower cut of 0.5~mas to exclude the central bright pixel of the simulation images.

With a radial cut of 4~mas, the contrast is $5 \pm 1 \%$ for the November epoch, $6 \pm 1 \%$ for the December epoch, 16\% for the original simulation image, and 12\% for the degraded resolution image (with negligible error bars). We also estimated the contrast over a sample of 30 randomly chosen snapshots of the simulation. We obtained an average value of $23 \pm 1~\%$ for the original images and $16 \pm 1~\%$ for the degraded resolution images. It appears that CE Tau presents a lower contrast than the 3D RHD simulation, and that the best matching snapshot corresponds to a lower contrast configuration of the convective pattern. This may come from the broadband imaging we did with \texttt{SQUEEZE} by merging all the spectral channels. However, this flattening of the convective pattern should also be observable on the 3D simulation. Therefore, we suggest that this less contrasted convective pattern is real. This can be  a consequence of the younger evolutionary stage of CE Tau compared for example with Antares (Sect. \ref{Sect:Stellar_params} and \citeads{2013A&A...555A..24O}) or to its stellar parameters, which  are quite different from those involved in the convective simulation. In particular, CE Tau is warmer than the simulation and has a smaller stellar radius: its surface gravity is higher than the model. The combination of these differences can lead to a less prominent convective pattern than the numerical model. Producing these simulations requires a lot of computer resources and it is not yet  possible to tailor this simulation grid to the needs of the observations of the various individual RSG stars. It is also possible that CE Tau is now experiencing a quieter episode of convective activity. This was previously observed by \citetads{2015MNRAS.446.3277C}, who derived a centrosymmetry parameter that was higher than the best 3D RHD simulations at that time. Moreover, a change in the convective activity has already been observed on the prototypical RSG Betelgeuse \citepads{2016A&A...588A.130M,2017A&A...602L..10O,2018A&A...609A..67K}.

Additionally, we note that the contrast of the reconstructed images increases monotonically when we increase the radial cut value. This can be the consequences of inhomogeneities lying at the edge rather than at the center of the disk for the images. However, it can also be a consequence of a residual LD effect. For the simulation, the trend is different with a contrast decreasing between the center and the edge of the star before being subjected to the limb effect. This may indicate that the simulation produces more inhomogeneities at the center of the disk.

\section{Conclusion \label{Sect:Conclusion}}

Our VLTI/PIONIER observations, collected at two different epochs, allowed us to derive refined values of the angular diameter of CE Tau. Using archival photometric measurements we were able to derive updated basic stellar parameters for this star, as well as its mass and its age using evolutionary models. These values allowed us to pick up a 3D RHD simulation and the interferometric observations constrained the temporal snapshots and their rotation angle. We were able to compare the best intensity image of the simulation with reconstructed images for our two epochs of observations. The simulation presents a higher radial contrast variation than the images. We suggest that it may be a consequence of the lower effective temperature of the simulation compared with CE Tau. As the observations on individual RSG stars are becoming more and more precise and numerous, such models become mandatory in order to properly identify the convective activity of these stars. As previous interferometric observations showed a vigorous convective activity, we also suggest that CE Tau is undergoing a quieter convective episode.

As already mentioned for the RSG star Betelgeuse (see Sect. \ref{Sect:Intro}), \citetads{2016A&A...591A.119A} have shown that the presence  of bright spots at the stellar surface can be traced by  spectropolarimetric observations. We have collected contemporaneous (Nov. -- Dec. 2016) spectropolarimetric observations of CE Tau  during a large program on the TBL/Narval instrument.  Although the PIONIER and the Narval instruments do not look at the same height in the atmosphere, and so probably probe a different convective scale, the combination of the two techniques  can improve our understanding of the convective surface of the star.  Comparison between interferometric and spectropolarimetric observations of CE Tau for the period of November -- December 2016 will be presented in a forthcoming paper (Tessore et al. in prep).

Studying the full temporal evolution of the observed convection would require more than two months of coverage. However, these observations represent the basis of future temporal monitoring of the convective pattern of RSG stars. This information is important for assessing the physical recipe of numerical simulations that aim to understand the atmosphere of RSG stars and to explain the origin of their mass loss.

\begin{acknowledgements}
        We would like to thank the anonymous referee whose suggestions and comments led to improvements in this article.
        We are grateful to  ESO's Director-General, Prof. Tim de Zeeuw, for the allocation of observing time to our program, and to the Paranal Observatory team for the successful execution of the observations.
        This project has received funding from the European Union’s Horizon 2020 research and innovation program under the Marie Sk\l{}odowska-Curie Grant agreement No. 665501 with the research Foundation Flanders (FWO) ([PEGASUS]$^2$ Marie Curie fellowship 12U2717N awarded to M.M.).
        A.L. acknowledges financial support from ``Programme National de Physique Stellaire'' (PNPS) of CNRS/INSU, France.
        F.B. acknowledges funding from NSF awards 1445935 and 1616483.
        We acknowledge with thanks the variable star observations from the AAVSO International Database that were contributed by observers worldwide and used in this research.
        We used the SIMBAD and VIZIER databases at the CDS, Strasbourg (France)\footnote{Available at \url{http://cdsweb.u-strasbg.fr/}}, and NASA's Astrophysics Data System Bibliographic Services.
        This research has made use of the Jean-Marie Mariotti Center's \texttt{Aspro}\footnote{Available at \url{http://www.jmmc.fr/aspro}} service, and of the \texttt{SearchCal} service\footnote{Available at \url{http://www.jmmc.fr/searchcal}} (co-developed by FIZEAU and LAOG/IPAG).
        This research made use of Matplotlib \citep{Hunter:2007}, Astropy\footnote{Available at \url{http://www.astropy.org/}}, a community-developed core Python package for Astronomy \citepads{2013A&A...558A..33A}, and  Uncertainties\footnote{Available at \url{http://pythonhosted.org/uncertainties/}}: a Python package for calculations with uncertainties.
\end{acknowledgements}

%__________________________________bliography
\bibliographystyle{aa}
\bibliography{./biblio}

\begin{appendix}

\section{Observation log}

%__________________________________Table of observations
\begin{table}[!ht]
        \caption{Log of the PIONIER observations of CE Tau and its calibrators. The $^*$ corresponds to the November dataset, the $^+$ to the December dataset (see Sect. \ref{Sect:Diameter}).}
        \label{Tab:ObsLog}
        \centering
        \begin{tabular}{llll}
                \hline \hline
                \noalign{\smallskip}
                Date & Time (UT) & Star & Configuration\\
                \hline
                \noalign{\smallskip}
                2016-Nov-14$^{*,+}$ & 03:45 & HR 1684 & A0-B2-C1-D0 \\
                            & 04:09 & CE Tau & A0-B2-C1-D0 \\
                            & 04:20 & $\phi$02 Tau & A0-B2-C1-D0 \\
                            & 04:28 & HR 1684 & A0-B2-C1-D0 \\
                            & 04:36 & CE Tau & A0-B2-C1-D0 \\
                            & 04:44 & $\phi$02 Tau & A0-B2-C1-D0 \\
                2016-Nov-22$^*$ & 05:07 & HR 1684 & D0-G2-J3-K0\\
                            & 05:16 & CE Tau & D0-G2-J3-K0\\
                            & 05:24 & HR 1684 & D0-G2-J3-K0\\
                            & 05:36 & CE Tau & D0-G2-J3-K0\\
                            & 05:46 & HR 1684 & D0-G2-J3-K0\\
                            & 05:56 & HR 1684 & D0-G2-J3-K0\\
                            & 06:19 & CE Tau & D0-G2-J3-K0\\
                            & 06:24 & HR 1684 & D0-G2-J3-K0\\
                            & 06:35 & CE Tau & D0-G2-J3-K0\\
                            & 06:45 & HR 1684 & D0-G2-J3-K0\\
            2016-Dec-22 & 03:06 & HR 1684 & D0-G2-J3-K0\\
                        & 03:42 & CE Tau & D0-G2-J3-K0\\
                        & 03:55 & $\phi$02 Ori & D0-G2-J3-K0\\
                        & 04:04 & CE Tau & D0-G2-J3-K0\\
                        & 04:25 & HR 1684 & D0-G2-J3-K0\\
            2016-Dec-23$^+$ & 02:34 & $\phi$02 Ori & D0-G2-J3-K0\\
                        & 03:28 & CE Tau & D0-G2-J3-K0\\
                        & 03:53 & HR 1684 & D0-G2-J3-K0\\
                        & 03:59 & CE Tau & D0-G2-J3-K0\\
                        & 04:16 & $\phi$02 Ori & D0-G2-J3-K0\\
                        & 04:26 & $\phi$02 Ori & D0-G2-J3-K0\\
                        & 04:35 & CE Tau & D0-G2-J3-K0\\
                        & 04:48 & $\phi$02 Ori & D0-G2-J3-K0\\
                        & 04:56 & CE Tau & D0-G2-J3-K0\\
                        & 05:09 & $\phi$02 Ori & D0-G2-J3-K0\\
                \hline
        \end{tabular}
        \tablefoot{Other executions of the observing blocks exist in the archive, but obtained a low quality grade.}
\end{table}

\section{$\chi^2$ maps for LDD and Gaussian spots models}

\begin{figure*}
        \centering
        \resizebox{\hsize}{!}{\includegraphics{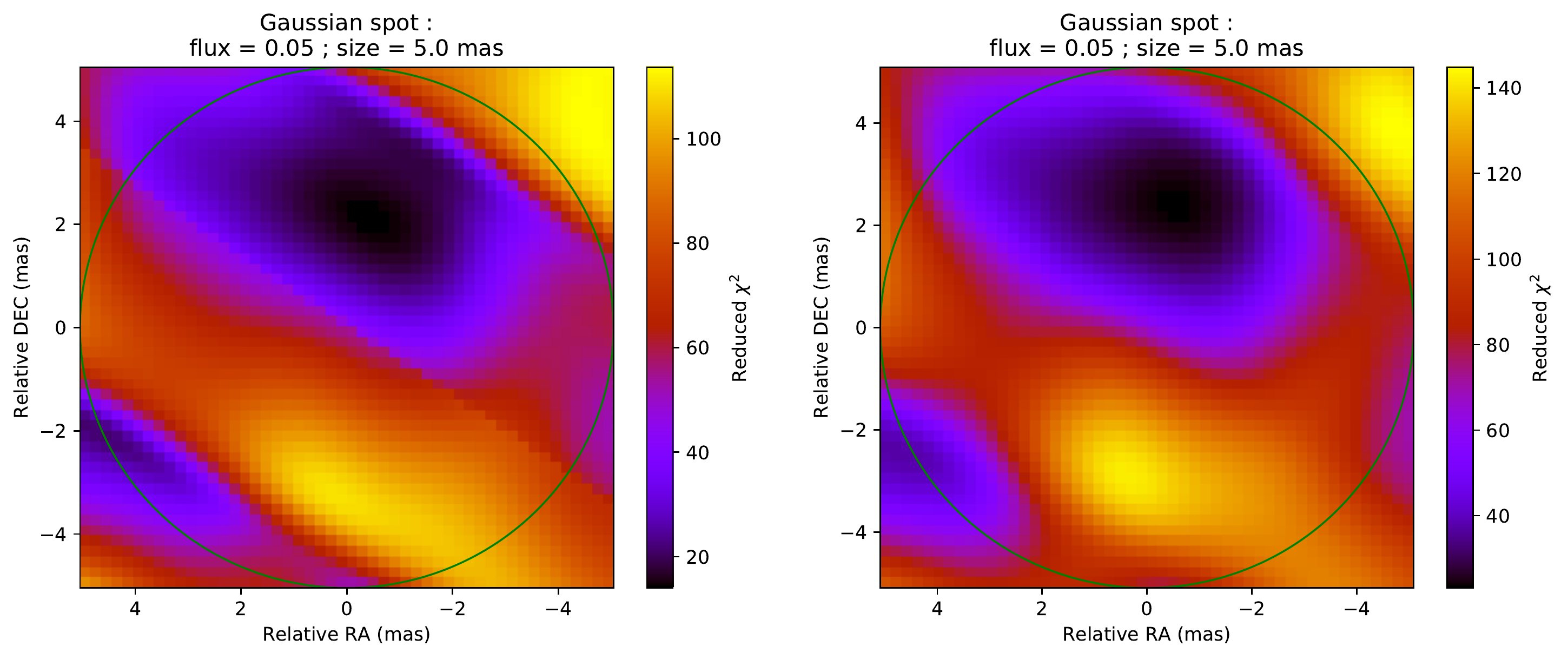}}
        \caption{$\chi^2$ maps for the single-spot model developed in Sect. \ref{Sect:spots}. \textit{Left:} November dataset. \textit{Right:} December dataset. \label{Fig:chi2map_1spot}}
\end{figure*}

%\begin{figure*}
%       \centering
%       \resizebox{\hsize}{!}{\includegraphics{chi2map_2spots}}
%       \caption{Same as Fig. \ref{Fig:chi2map_1spot} for the two spots model. \label{Fig:chi2map_2spots}}
%\end{figure*}

\section{Image reconstruction: additional figures}

\subsection{\texttt{SQUEEZE} reconstruction}

\begin{figure*}
        \centering
        \resizebox{\hsize}{!}{\includegraphics{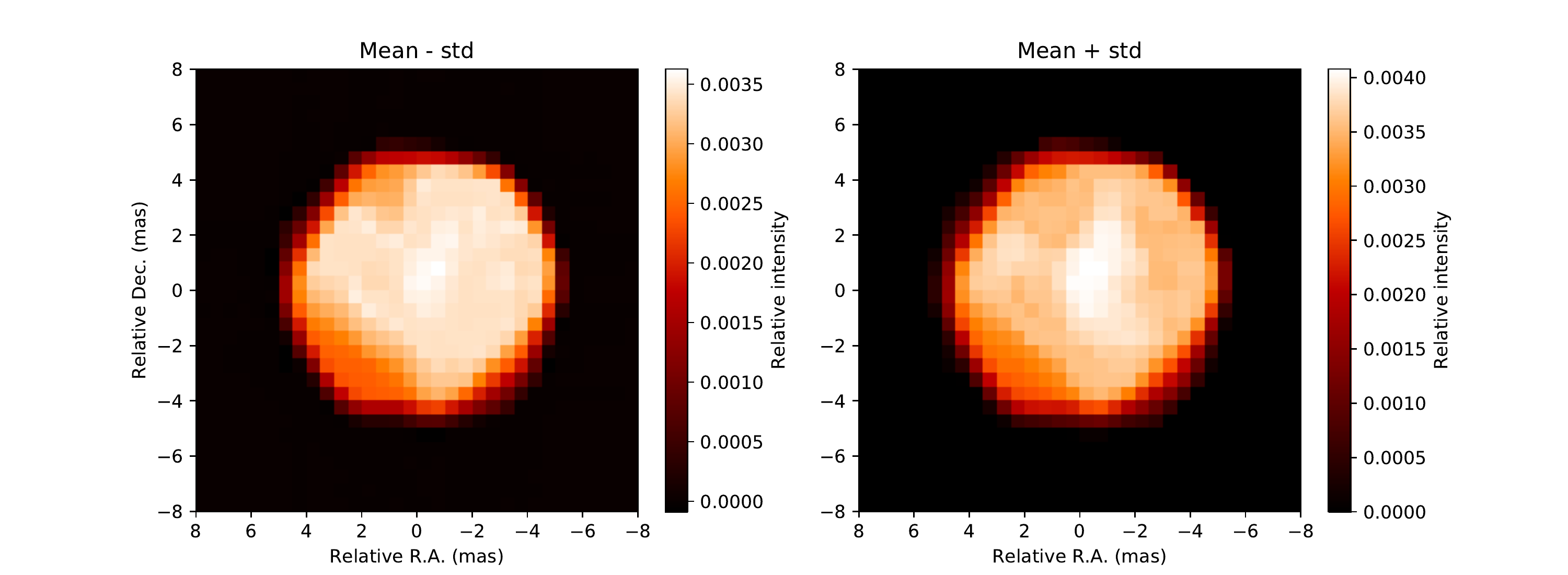}}
        \resizebox{\hsize}{!}{\includegraphics{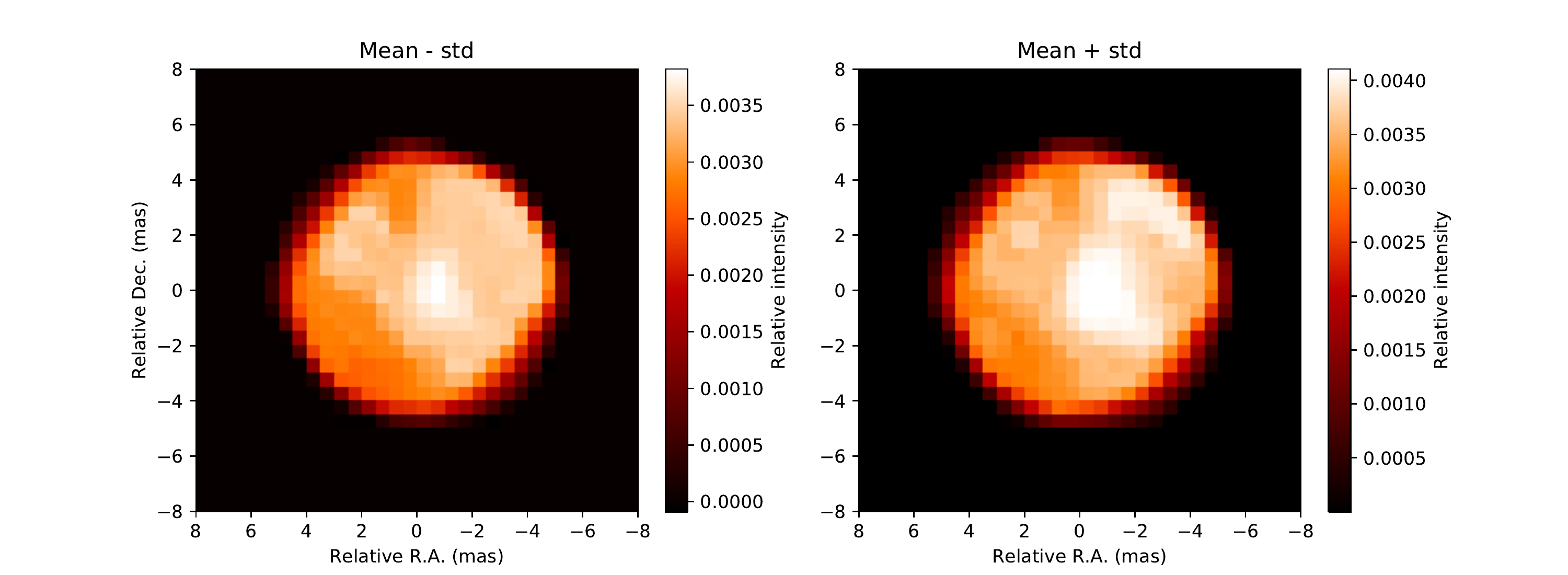}}
        \caption{Mean reconstructed image (see Sect. \ref{Sect:Image}) minus the standard deviation (\textit{left column}) and plus the standard deviation (\textit{right column}). The top row corresponds to the November dataset and the bottom row to the December dataset. \label{Fig:image_recons_std}}
\end{figure*}

In addition to the mean reconstructed images presented in Sect. \ref{Sect:Image}, we present in Fig. \ref{Fig:image_recons_std} the mean images plus and minus the standard deviation.

\subsection{\texttt{MIRA} reconstruction}

\begin{figure}
        \centering
        \resizebox{\hsize}{!}{\includegraphics{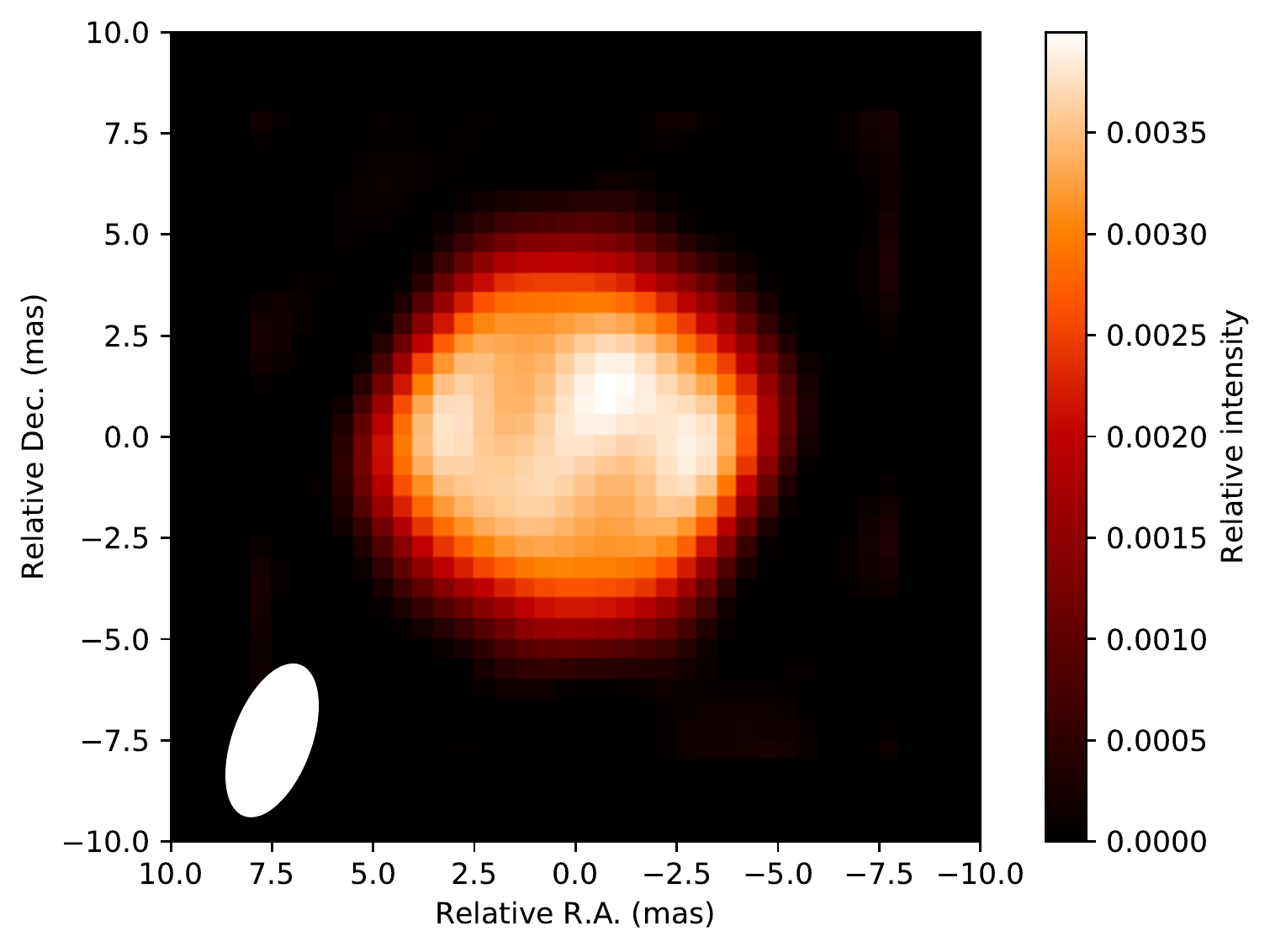}}
        \resizebox{\hsize}{!}{\includegraphics{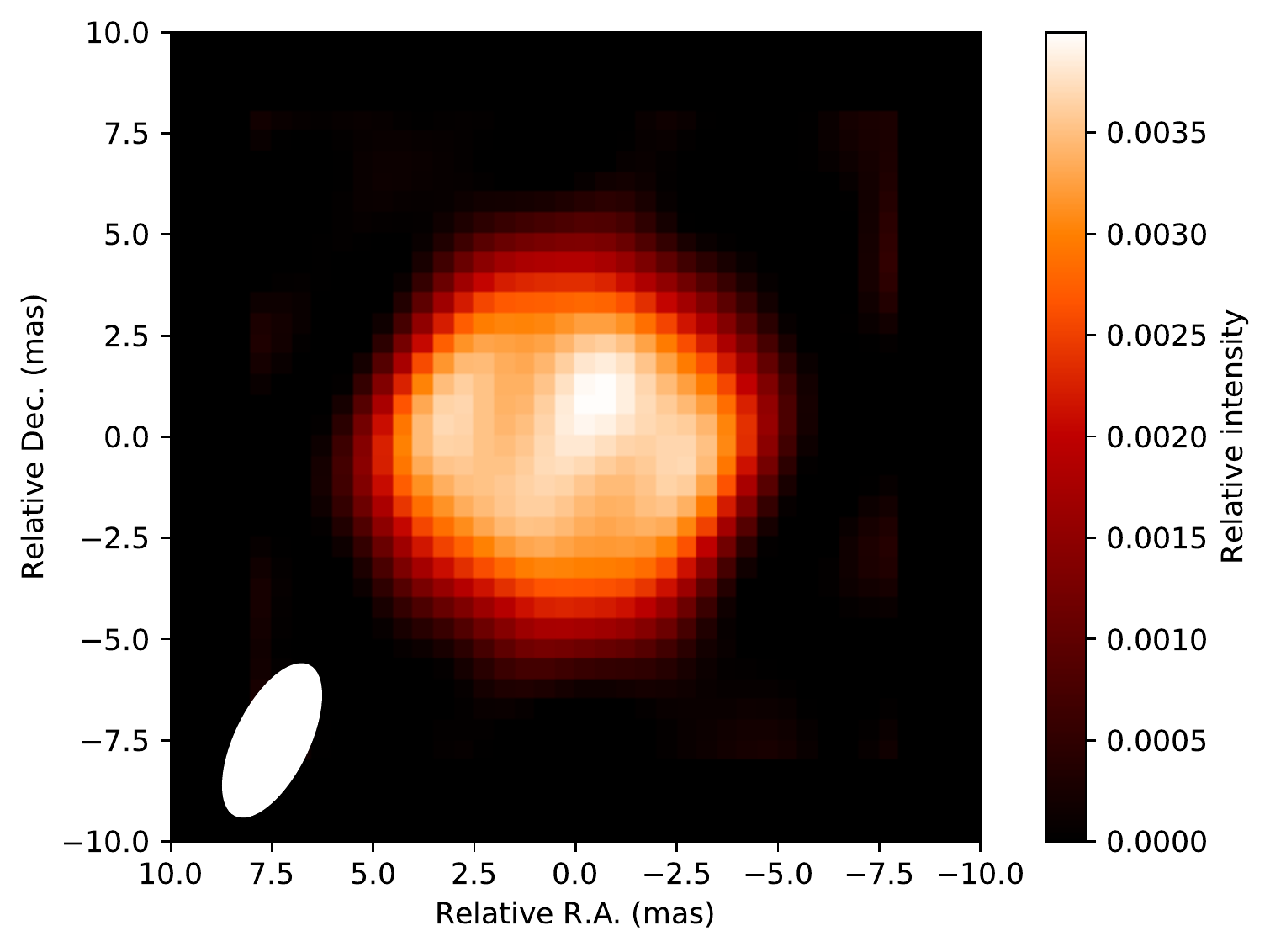}}
        \caption{\texttt{MIRA} reconstructed images of CE Tau at a 0.5~mas resolution. The November image is at the top and the December image at the bottom.\label{Fig:Mira}}
\end{figure}

Figure \ref{Fig:Mira} shows the reconstructed image of CE Tau obtained with \texttt{MIRA}. For more details on the reconstruction process with this algorithm, see Sect. \ref{Sect:Image}.

\end{appendix}

\end{document}